\documentclass[preprintnumbers,prd,twocolumn,showpacs,floatfix,preprintnumbers,superscriptaddress,nofootinbib]{revtex4-2}

\usepackage{graphicx}
\usepackage{epsfig}
\usepackage{bm}
\usepackage{amssymb}
\usepackage{float}
\usepackage{amsmath}
\usepackage{subfigure}
\usepackage{dcolumn}
\usepackage[colorlinks]{hyperref}
\usepackage[usenames,dvipsnames]{color}
\hypersetup{
     breaklinks=true,
    pdfstartview={FitH},    
    colorlinks=true,       
    linkcolor=blue,          
    citecolor=red,        
    filecolor=magenta,      
    urlcolor=blue,           
    anchorcolor=green,      
    linktocpage=true
}
\usepackage{orcidlink}

\newcommand{\Mpl}{M_{\textrm{Pl}}}
\renewcommand{\(}{\left(}
\renewcommand{\)}{\right)}

\def\al{\alpha}

\def\Om{\Omega}
\def\sig{\sigma}

\def\lam{\lambda}

\def\S{\mathcal{S}}

\def\doi{http://doi.org}

\def\r{\mathrm{r}}
\def\g{\mathrm{g}}

\def\m{\mathrm{m}}

\def\d{\mathrm{d}}

\allowdisplaybreaks

\begin{document}

\title{A comparison between axion-like and power law potentials in cosmological background}

\author{Md. Wali Hossain\orcidlink{0000-0001-6969-8716}}
\email{mhossain@jmi.ac.in}
\affiliation{Department of Physics, Jamia Millia Islamia, New Delhi, 110025, India}

\author{Afaq Maqsood}
\email{afaq2206145@st.jmi.ac.in}
\affiliation{Department of Physics, Jamia Millia Islamia, New Delhi, 110025, India}


\begin{abstract}
In this paper, we compare the scalar field dynamics in axion-like and power law potentials for both positive and negative values of the exponents. We find that, for positive exponents, both the potentials exhibit similar scalar field dynamics and it can be difficult to distinguish them at least at the background level. Even though the potentials are oscillatory in nature for positive exponents scaling solutions can be achieved for larger values of the exponent for which the dynamics can be different during early times. Because of the presence of this scaling nature there is a turnaround in the values of the scalar field equation of state as we increase the values of the exponent in both the potentials. This indicates the deviation from the oscillatory behaviour for the larger values of the exponent. For negative values of the exponent, the dynamics of the scalar field is distinguishable and axion-like potential can give rise to cosmologically viable tracker solutions unlike the power law potentials. For negative values of the exponent, axion-like potential can behave like a cosmological constant around its minima and the dark energy scale can be related to the potential scale. Due to the cosmological constant like behavior of the axion-like potential for negative exponent around its minima the late time dynamics can be similar to $\Lambda$CDM and we get similar observational constraint on the parameters for both $\Lambda$CDM and axion-like potential with negative exponent. So, while for positive exponents we may not distinguish the two potentials for negative exponents the dynamics of the scalar field is distinguishable. 
\end{abstract}

\maketitle
\section{Introduction}
\label{sec:intro}
Recent cosmological observations suggest that our
present universe is expanding with an acceleration \cite{SupernovaSearchTeam:1998fmf,SupernovaCosmologyProject:1998vns,Planck:2018vyg,Brout:2022vxf}. The reason of the late time acceleration still remains a mystery but the cosmological constant (CC), $\Lambda$, has appeared as the simplest yet the most viable explanation especially by the observations of cosmic microwave background (CMB) radiation\cite{Planck:2013pxb,Planck:2018vyg}. However, the CC is plagued with two issues, namely the fine-tuning
problem \cite{Martin:2012bt} and the cosmic coincidence problem \cite{Zlatev:1998tr,Steinhardt:1999nw}. This motivates us to look for other possible solutions, {\it e.g.}, making the dark energy dynamical \cite{Copeland:2006wr,Sahni:1999gb} or modifying the gravity \cite{Clifton:2011jh}. In this paper, we study the previous case. Considering a scalar field instead of CC makes the equation of state (EoS) dynamical. We study the cosmological dynamics of scalar field for two specific potentials, axion-like \cite{Marsh:2015xka,Hlozek:2014lca,Poulin:2018cxd,Poulin:2018dzj} and power law potentials \cite{Ratra:1987rm}.

Slowly rolling scalar fields, known as the quintessence field \cite{Ratra:1987rm,Wetterich:1987fk,Wetterich:1987fm}, can also explain the late time acceleration \cite{Copeland:2006wr}. Scalar fields can also solve the cosmic coincidence problem for some
specific scenarios known as the tracker models \cite{Zlatev:1998tr,Steinhardt:1999nw} in which the scalar field energy density tracks the
background energy density in the past and takes over matter
during the recent past. A power law potential can give rise to tracker solutions for larger negative values of the exponent \cite{Zlatev:1998tr,Steinhardt:1999nw}. But the problem with this scenario is that the EoS of the scalar field can be much greater than $-1$ at present for some potentials like power law potentials \cite{Steinhardt:1999nw}. But the advantage is that the late time solution is independent of a wide range of initial conditions as it is an attractor solution which solves the coincidence problem. There is another class of scalar field models known as the thawing models \cite{Caldwell:2005tm,Scherrer:2007pu}. In this scenario the scalar field behaves as the CC in the past and starts evolving from the recent past which ensures the difference with the CC. Apart from these two dynamics we can also have scaling behaviour for steep exponential potential in which the scalar field scales the background energy density \cite{Copeland:1997et}. It's an attractor solution, so the scalar field scales forever in this dynamics. To exit from attractor scaling behaviour we need to modify the potential, {\it e.g.}, double exponential potential \cite{Barreiro:1999zs} or cosh potential \cite{Sahni:1999qe} can give scaling behaviour followed by late time acceleration. This dynamical behaviour is generally known as the scaling-freezing dynamics.

Even though the CC is the most suitable explanation of late time acceleration recent tensions in cosmology \cite{Perivolaropoulos:2021jda} {\it e.g.}, tension in the measurements of present value of Hubble parameter ($H_0$) \cite{Riess:2021jrx,Kamionkowski:2022pkx} and the tension in the growth rate known as the $S_8=\sig_8 \sqrt{\Om_{\m0}/0.3}$ tension \cite{Perivolaropoulos:2021jda,Kilo-DegreeSurvey:2023gfr}, where $\sig_8$ is the standard deviation of
matter density fluctuations at present for linear perturbation in spheres of radius $8h^{-1}{\rm Mpc}$ and $\Om_{\m0}$ is the present value of matter density parameter, raise the question about the validity of the standard $\Lambda$CDM model. The $\sim 5\sig$ tension between the local measurement of $H_0$ by the SH0ES team ($H_0=73.04\pm1.04~{\rm km~sec^{-1}~Mpc^{-1}}$) \cite{Riess:2021jrx} and its constraint coming from the observation of cosmic microwave background (CMB) radiation assuming the standard $\Lambda$CDM model ($H_0=67.4\pm 0.5~{\rm km~sec^{-1}~Mpc^{-1}}$) \cite{Planck:2018vyg} has opened a new challenge in cosmology. The discrepancy between the two measurements can be due to systematic errors or some new physics is there \cite{Kamionkowski:2022pkx,Freedman:2023jcz,Bernal:2016gxb,Knox:2019rjx,Verde:2023lmm}. To explain the tension with new physics it is very clear that we have to go beyond $\Lambda$CDM. In this regard, scalar fields can play an interesting roll as we can evade the tension, at least to some extent, by modifying the cosmic history either during early times or late times \cite{Kamionkowski:2022pkx,Knox:2019rjx,DiValentino:2021izs,Hu:2023jqc}.

Modifying the early expansion of the universe can be done by early dark energy (EDE) scenario \cite{Poulin:2018cxd,Poulin:2018dzj,Poulin:2023lkg,Smith:2019ihp,Agrawal:2019lmo} where the scalar field density parameter ($\Omega_\phi$) has a small but finite value ($<0.1$) during the matter radiation equality. Among the different EDE models \cite{Poulin:2023lkg} two scenarios have gained a lot of interest, one is the axion-like potential $\sim (1-\cos \phi)^n$ \cite{Marsh:2015xka,Poulin:2018cxd,Poulin:2018dzj,Poulin:2023lkg,Smith:2019ihp,Hlozek:2014lca} and the another one is the power law potential $\sim \phi^{2n}$ \cite{Ratra:1987rm,Agrawal:2019lmo}, where $n$ is a constant. The axion-like potential has minima and around the minima the potential behaves like $\phi^{2n}$, {\it i.e.}, power law potential for positive values of $n$ . So, Axion-like potential has power law behaviour around the minima which should give similar dynamics of the scalar field during the late time when the scalar field rolls near the minima for positive values of $n$. The possible deviation in oscillations in the two potentials have been studied in \cite{Norton:2020ert}. Now, the question remains whether the scalar field dynamics is different in the two potentials during the early times when the scalar field is away from the minima.
And if it is, then is it significant? We address these issues in this paper.  We also explore the dynamics for negative values of $n$. Even though the axion-like potential reduces to a power law potential around the minima for positive values of $n$ the dynamics of the scalar field can be different for negative values of $n$ as the axion-like potential still has a minima but the power law is a runaway potential. This difference in the nature of the potentials can lead to distinguishable features in the cosmological dynamics and therefor we examine this in detail. So, in this paper, our aim is to compare the cosmological dynamics in two widely considered potentials and try to see if there is any degeneracy in the dynamics. In this regard, we also want to explore the effects of the functional form and the values of the functions $\lam=V_\phi(\phi)/V(\phi)$ and $\Gamma=V_{\phi\phi}(\phi)V(\phi)/V_\phi(\phi)^2$, where $V(\phi)$ is the potential and the subscript $\phi$ denotes the derivative with respect to (w.r.t.) $\phi$, on the scalar field dynamics. So, in other words, we want to see whether the dynamics of the scalar field in the cosmological background depends on the nature of the potential solely or the values and forms of the functions $\lam$ and $\Gamma$ mainly decides the the dynamics irrespective of the explicit functional form of the potential. To study this we consider two potentials, axion-like and power law, phenomenologically as examples. To compare the dynamics in the two potentials we first study the fixed points and their stability which turns out to be similar and does not give us any distinguishable information. So, we analyse each scenario numerically. We also compare the EDE solutions in both the potentials and we argue that the EDE solutions are almost identical and we don't expect anything distinguishable.

We introduce the scenario with the background cosmological equations in Sec.~\ref{sec:model}. In Sec.~\ref{sec:dyn} we study the stability of the fixed points by analysing the dynamical system. The comparison has been done numerically in Sec.~\ref{sec:num} in which we first study the nature of potentials in Sub Sec.~\ref{sec:nature}. Then we study the dynamics for positive values of $n$ in Sub Sec.~\ref{sec:posn}. The study for negative values of $n$ has been done in Sub Sec.~\ref{sec:negn}. The comparison in EDE scenario for both the potentials has been done in Sec.~\ref{sec:EDE}. In Sec.~\ref{sec:obs} observational constraints on the model parameters for tracker models in axion-like potential has been studied. We finally summarise and conclude in Sec.~\ref{sec:conc}.

\section{The Model}
\label{sec:model}
We consider the following action of a minimally coupled canonical scalar field
\begin{align}
\S=\int \d^4x\sqrt{-\g}\Bigl [\frac{\Mpl^2}{2} R-\frac{1}{2}\partial_\mu\phi\partial^\mu\phi - V(\phi) \Bigr]+ \S_{\rm B}\, ,
\label{eq:action}
\end{align}
where $\Mpl=1/\sqrt{8\pi G}$ is the reduced Planck mass, $V(\phi)$ is the potential of the field and $\S_{\rm B}=\S_\m+\S_\r$ with $\S_\m$ and $\S_\r$ being the actions for matter and radiation respectively.

Variation of the action (\ref{eq:action}) w.r.t. the metric $g_{\mu\nu}$ gives the Einstein's field equation
\begin{align}
\Mpl^2 G_{\mu\nu}= T_{(\m)\mu\nu}+T_{(\r)\mu\nu}+T_{(\phi)\mu\nu} \, ,
\label{eq:ee}
\end{align}
where
\begin{align}
T_{(\phi)\mu\nu}=& \phi_{;\mu}\phi_{;\nu}-\frac{1}{2}\g_{\mu\nu}(\nabla\phi)^2 -\g_{\mu\nu}V(\phi) \, ,
\label{eq:emt_phi}
\end{align}
is the energy-momentum tensor of the scalar field. Variation w.r.t. the scalar field $\phi$ gives the equation of motion of the scalar field 
\begin{align}
& \Box \phi-\frac{\d V}{\d\phi}=0 \, ,
\label{eq:eom_phi}
\end{align}

In flat  Friedmann-Lema\^itre-Robertson-Walker (FLRW) geometry, with metric
\begin{align}
\label{metric0}
 ds^2 = -dt^2 +a(t)^2\delta_{ij} dx^idx^j ~,
\end{align}
where $a(t)$ is the scale factor, the Friedman equations are given by
\begin{eqnarray}
 3H^2\Mpl^2 &=& \rho_\m+\rho_\r+\frac{1}{2}\dot\phi^2+V(\phi) \,  
 \label{eq:Friedmann1}\\
 \(2\dot H+3H^2\)\Mpl^2&=&-\frac{1}{3}\rho_\r-\frac{1}{2}\dot\phi^2+V(\phi) \, ,
 \label{eq:Friedmann2}
\end{eqnarray}
where, $H=\dot a/a$, is the Hubble parameter with {\it dot} representing the derivative w.r.t. time. $\rho_\m$ and $\rho_\r$ are the matter and radiation energy densities respectively. The energy density and pressure of the scalar field are given by
\begin{eqnarray}
    \rho_\phi &=& \frac{1}{2}\dot\phi^2+V(\phi) \, , 
    \label{eq:rhophi}\\
    p_\phi &=& \frac{1}{2}\dot\phi^2-V(\phi) \, .
    \label{eq:pphi}
\end{eqnarray}
The equation of motion of the scalar field has the standard
form
\begin{equation}
 \ddot\phi+3H\dot\phi+\frac{\d V}{\d\phi}=0 \, .
 \label{eq:eomphi}
 \end{equation}

In this paper we are going to study the background cosmological dynamics of the scalar field for an axion-like potential given by \cite{Marsh:2015xka,Poulin:2018cxd,Poulin:2018dzj,Smith:2019ihp}
\begin{equation}
    V(\phi)=V_0\(1-\cos\(\frac{\phi}{f_{\rm pl}}\)\)^n \, , 
    \label{eq:pot}
\end{equation}
where $V_0$, $f_{\rm pl}$ and $n$ are constants, and compare the same for the power law potential \cite{Ratra:1987rm}
\begin{equation}
    V(\phi)=V_0 \(\frac{\phi}{f_{\rm pl}}\)^{2n} \, .
    \label{eq:pot_pl}
\end{equation}

The axion-like potential~\eqref{eq:pot} is periodic in nature which repeats itself after one full cycle or time period. So, by any simple transformation in the scalar field we can go from one cycle to any cycle which does not affect the dynamics of the scalar field. That means, we can study the dynamics of the scalar field in one cycle as the dynamics will be same in each full cycle. Also, as a periodic function the axion-like potential has maxima and minima and around the minima the axion-like potential reduces to power law potential. In the following discussion of this paper we compare the cosmological dynamics of the scalar field in these two potentials for both positive and negative values of the parameter $n$.

\section{Dynamical system analysis}
\label{sec:dyn}
 
To compare the dynamics of the scalar field in the axion-like~\eqref{eq:pot} and power law~\eqref{eq:pot_pl} potentials we first study the dynamical system for the axion-like potential and analyse the nature of stability of the fixed points. This gives us the idea about the stability of the background cosmological solutions for the axion-like potential (for power law potential see Ref.~\cite{Bahamonde:2017ize}). So, in this section, we examine whether we get any specific fixed points which can distinguish the dynamics of the scalar field for the axion-like potential from the power law potential. To form the dynamical system we define the following dimensionless variables
\begin{eqnarray}
    x &=&\frac{\dot{\phi}}{\sqrt6 H\Mpl} \; , 
    \label{eq:x}\\
    y &=&\frac{\sqrt V}{\sqrt3 H \Mpl} \; , 
    \label{eq:y}\\
    \Omega_\r &=&\frac {\rho_r}{3 H^2\Mpl^2} \ ; ,
    \label{eq:omr}\\
    \lambda &=& -\Mpl\frac{V_\phi(\phi)}{V} \; , 
    \label{eq:lam}\\
    \Gamma &=& \frac{V_{\phi\phi}(\phi)V(\phi)}{V_\phi(\phi)^2} \; .
    \label{eq:gam}
\end{eqnarray}

While the variable $\lam$ signifies the slope of the potential the variable $\Gamma$ represents the nature of the potential, {\it e.g.}, $\Gamma=1$ for an exponential potential of constant slope $\lam$. For the axion-like potential~\eqref{eq:pot},
\begin{eqnarray}
    \lam &=& -\frac{n}{f}\cot\(\frac{\phi}{2f_{\rm pl}}\) \, ,
    \label{eq:lam0}\\
    \Gamma &=& 1-\frac{1}{2n}-\frac{n}{2f^2\lambda^2} \, ,
    \label{eq:Gamma}
\end{eqnarray}
where, $f=f_{\rm pl}/\Mpl$. For power law potential~\eqref{eq:pot_pl}
\begin{eqnarray}
    \lam &=& -\frac{2n}{(\phi/\Mpl)} \, ,
    \label{eq:lam0_pl}\\
    \Gamma &=& 1-\frac{1}{2n}\, .
    \label{eq:Gamma_pl}
\end{eqnarray}
As the function $\Gamma$ represents the nature of the potential and from Eqs.~\eqref{eq:Gamma} and \eqref{eq:Gamma_pl} we can see that the difference in the functional form of $\Gamma$ for the two potentials is the last term of Eq.~\eqref{eq:Gamma}. This term can be negligible for large values of the slope $\lam$ which makes both the $\Gamma$s identical. So, the scalar field dynamics is expected to be similar for both the potentials though it will also depend on the values of $\lam$. Now it would also be interesting to see whether the last term of Eq.~\eqref{eq:Gamma} makes any significant difference in the dynamics of the scalar field. In this section we look for these answers by doing fixed points analysis. In the next section we discuss the effects of the nature and values of the functions $\lam$ and $\Gamma$ on the dynamics of the scalar field for both the potentials.

From Eq.~\eqref{eq:Friedmann1} we see that the dimensionless variables follow the constraint equation $x^2+y^2+\Om_\r+\Om_\m=1$, where $\Om_\m=\rho_m/3H^2\Mpl^2$ is the density parameter of matter. The density parameter of the scalar field is defined as $\Om_\phi=x^2+y^2$ while $\Om_\r$ is that for radiation. Effective EoS and the EoS of the scalar field, in terms of the dimensionless variables, are given by
\begin{eqnarray}
w_{\text{eff}}&=& -\(1+\frac{2}{3}\frac{\dot H}{H^2}\) = x^2-y^2+\frac{1}{3}\Om_\r \; , 
\label{eq:weff}\\
w_{\phi} &=& \frac{\frac{1}{2}\dot\phi^2-V(\phi)}{\frac{1}{2}\dot\phi^2+V(\phi)} = \frac{x^2-y^2}{x^2+y^2} \; .
\label{eq:wphi}
\end{eqnarray}

\begin{table*}[t]
\begin{center}
\caption{Fixed points of the autonomous system~\eqref{eq:xp2}-\eqref{eq:lamp2} with their nature of stability and the values of different cosmological parameters at the fixed points are listed.}
\label{tab:fp}
\resizebox{\textwidth}{!}{%
\begin{tabular}{c|c|c|c|c|c|c|c|c|c|c|c|c}\hline \hline
Point & $x$ & $y$ & $\Om_\r$ & $\xi$ & Eigenvalues & Existence  & Hyperbolicity & Stability & $\Omega_\m$ & $\Omega_\phi$ & $w_\phi$ & $w_{\rm eff}$ \\ 
\hline\hline
A & 0 & 0 & 1 & Any & 0, $2(1-\xi)$, $1-\xi$,  & All $n$ & Non-hyperbolic & Saddle for $\xi\neq 1$ & 0 & 0 & Indeterminate & 1/3 \\ 
&&&&& $-(1-\xi)$ &&&&&&& \\ \hline
B & 0 & 0 & 0 & Any & 0, $\frac{3}{2}(1-\xi)$, $-(1-\xi)$, & All $n$ & Non-hyperbolic & Saddle for $\xi\neq 1$ & 1 & 0 & Indeterminate & 0 \\ 
&&&&& $-\frac{3}{2}(1-\xi)$ &&&&&&& \\ \hline

C & $x$ & 0 & $\Om_\r$ & 1 & 0, 0, $-\frac{1}{n}\sqrt{\frac{3}{2}}x$, & Finite $n$ & Non-hyperbolic & Stable for $n>0$  & $1-x^2-\Om_\r$ & $x^2$ & 1 & $x^2+\frac{1}{3}\Om_\r$ \\ 
&&&&& $-\sqrt{\frac{3}{2}}x$ &&& 
and $x>0$ (Fig.~\ref{fig:yxi}) &&&& \\ 
&&&&&  &&& 
Saddle for $x<0$ &&&& \\ 
&&&&& &&& 
or $n<0$ and $x>0$ &&&& \\ 
&&&&& &&& 
Unstable for $n>0$ &&&& \\ 
&&&&& &&& 
~~~~~~~~ and $x<0$ &&&& \\ \hline

$D_\pm$ & 0 & $\pm 1$ & 0 & 0 &-4, -3, $-\frac{3}{2}(1+\sqrt{1+\frac{2}{3}n})$, & All $n$ & Hyperbolic & Stable for $n<0$ & 0 & 1 & -1 & -1 \\
&&&&& $-\frac{3}{2}(1-\sqrt{1+\frac{2}{3}n})$ &&& 
Saddle for $n>0$ &&&& \\
&&&&&  &&& 
(Fig.~\ref{fig:xix}) &&&& \\
\hline
\hline
\end{tabular}}
\end{center}
\end{table*}

Evolution of the universe is represented by the equations~\eqref{eq:Friedmann2} and \eqref{eq:eomphi}, which can be recast in terms of the evolution equations of the dimensionless variables~\eqref{eq:x}-\eqref{eq:lam} by the following autonomous system,
\begin{eqnarray}
\frac{dx}{dN} &=& -\frac{3}{2} x (1-x^2+y^2-\frac{1}{3}\Omega_r)+\sqrt{\frac{3}{2}} y^2\lam \; , 
\label{eq:xp}\\
\frac{dy}{dN} &=&\frac{3}{2} y  (1+x^2-y^2+\frac{1}{3}\Omega_r) - \sqrt{\frac{3}{2}} xy\lam \; , 
\label{eq:yp}\\
\frac{d\Omega_\r}{dN} &=&-\Omega_\r (1-3x^2+3y^2-\Omega_\r) \; , \\
\frac{d\lambda}{dN} &=&\sqrt6 x \lambda^2 (1- \Gamma)=\sqrt\frac{3}{2}x
\left( \frac{\lambda^2}{n}
+\frac{n}{f^2} \right) \; .
\label{eq:lamp}
\end{eqnarray}
where $N=\ln a$. Here we should mention that the autonomous system~\eqref{eq:xp}-\eqref{eq:lamp} is valid for $n\neq 0$ as apparently it seems that the Eq.~\eqref{eq:lamp} diverges as $n\to 0$. For $n=0$ the potential~\eqref{eq:pot} reduces to a constant potential {\it i.e.}, $\lam=0$. So, the autonomous system will not contain the $\lam$ variable for $n=0$.

As we have already constructed the autonomous system we can calculate the fixed points by equating the equations~\eqref{eq:xp}-\eqref{eq:lamp} to zero. Before that, we should notice from the Eq.~\eqref{eq:lam0} that $0\leq|\lam|<\infty$, and for $\lam\to\infty$ the Eq.~\eqref{eq:lamp} diverges. So the autonomous system~\eqref{eq:xp}-\eqref{eq:lamp} can give us the fixed points as long as $\lam$ is either zero or finite. To avoid this divergence problem we define a new variable corresponding to $\lam$ as \cite{Ng:2001hs,Bahamonde:2017ize} 
\begin{eqnarray}
    \xi=\lam/(1+\lam)
\end{eqnarray}
and rewrite the autonomous system. We can see that $0\leq\xi\leq1$ for $0\leq\lam\leq\infty$ {\it i.e.}, $\xi$ remains finite even if $\lam\to\infty$.

In terms of $\xi$ the last terms of Eqs.~\eqref{eq:xp} and \eqref{eq:yp} will have factor $\xi/(1-\xi)$ which blows up for $\xi\to 1$. To avoid this divergence in the autonomous system we use the transformation $\d N\to (1-\xi) \d N$ \cite{Bahamonde:2017ize} which does not change the dynamical behaviour of the system as there will be an overall multiplication of the term $(1-\xi)$ in the autonomous system but makes the dynamical system consistent around $\xi=1$ \cite{Bahamonde:2017ize}. So the final dynamical system, consistent around $\xi=1$, becomes
\begin{eqnarray}
\frac{dx}{dN} &=& -\frac{3}{2} x (1-\xi) (1-x^2+y^2-\frac{1}{3}\Omega_r)+\sqrt{\frac{3}{2}} y^2\xi \; ,~~~~ 
\label{eq:xp2}\\
\frac{dy}{dN} &=&\frac{3}{2} y (1-\xi)  (1+x^2-y^2+\frac{1}{3}\Omega_r) - \sqrt{\frac{3}{2}} xy\xi \; , 
\label{eq:yp2}\\
\frac{d\Omega_\r}{dN} &=&-\Omega_\r (1-\xi)(1-3x^2+3y^2-\Omega_\r) \; , \\
\frac{d\xi}{dN} &=&\sqrt\frac{3}{2}x (1-\xi)
\left( \frac{\xi^2}{n}
+\frac{n(1-\xi)^2}{f^2} \right) \; .
\label{eq:lamp2}
\end{eqnarray}

\begin{figure}[h]
\centering
\includegraphics[scale=.9]{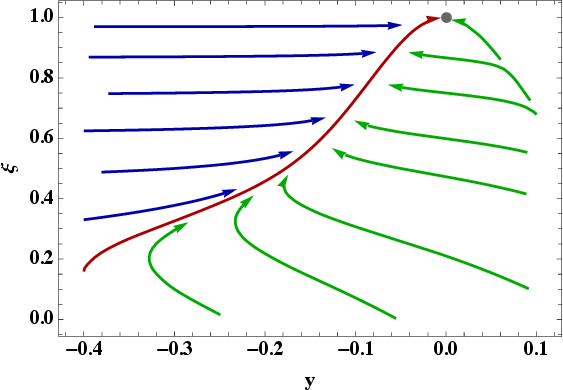}
\caption{Nature of stability of the point $C$ for $n,x>0$ in the $y-\xi$ plane. The gray dot represents the point $C$ {\it i.e.}, $(y,\xi)\equiv(0,1)$. We have considered $n=2$, $x=0.02$ and $\Om_\r=0$.} 
\label{fig:yxi}
\end{figure}

\begin{figure}[h]
\centering
\includegraphics[scale=.89]{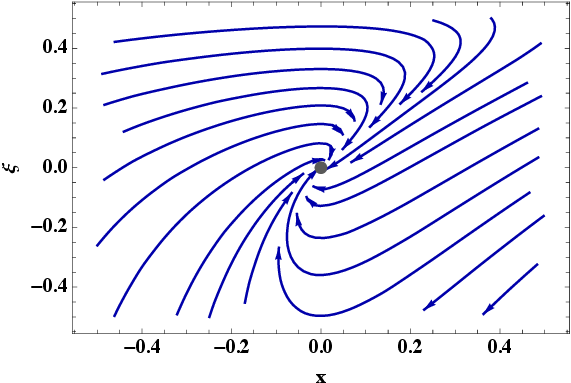} \vskip15pt
\includegraphics[scale=.89]{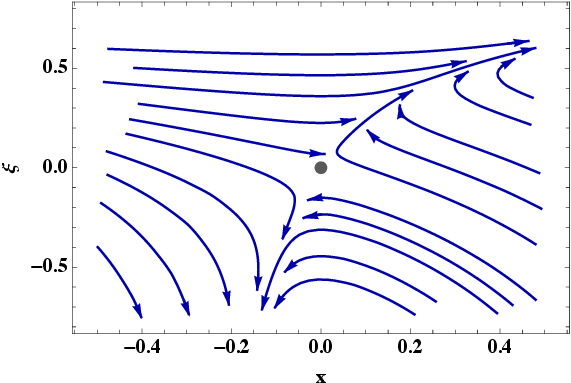}
\caption{Nature of stability of the point $D$ for $n=-2$ (upper) and $n=2$ (lower) has been shown in the $x-\xi$ plane. The gray dot represents the the points $D_\pm$ {\it i.e.}, $(x,\xi)\equiv(0,0)$. We have considered $y=1$. Here we have also shown the region $\xi>1$ to show the nature of the trajectories properly.} 
\label{fig:xix}
\end{figure}

The fixed points of the autonomous system~\eqref{eq:xp2}-\eqref{eq:lamp2} with their nature of stability are listed in the Tab.~\ref{tab:fp}. We can see that the fixed points and their stability are similar same as the power law potential~\eqref{eq:pot_pl} \cite{Bahamonde:2017ize}. The difference comes in the points $D_\pm$ as for the power law potential these points are non-hyperbolic but for the potential~\eqref{eq:pot} we have hyperbolic points with same condition on $n$ for the nature of stability. From Tab.~\ref{tab:fp} we can see that in the cosmological dynamics of axion-like potential we can have three stable points same as power law potential \cite{Bahamonde:2017ize}. Point $C$ is stable for $n,x>0$. We have shown this in Fig.~\ref{fig:yxi} in the $y-\xi$ plane by considering fixed values of $x$ and $\Om_\r$. We can see that if we perturb the system from the fixed point $(0,1)$ in $y-\xi$ plane then all the trajectories are moving towards the fixed point $C$ for $n,x>0$ which ensures the stable nature of the point $C$. Similarly, in the upper figure of Fig.~\ref{fig:xix} we have shown the stable nature of the fixed points $D_\pm$ for $n=-2<0$ in the $x-\xi$ plane. In this figure all the trajectories are moving towards the fixed points $D_\pm$. On the other hand, the lower figure of Fig.~\ref{fig:xix} shows that some of the trajectories are moving towards the fixed points $D_\pm$ and some of them are going away from it for $n=2>0$ in the $x-\xi$ plane. This shows the saddle nature of the fixed points $D_\pm$ for positive values of $n$.

From Tab.~\ref{tab:fp} we can see, from the fixed point analysis, only $D_\pm$ points can give rise to late time acceleration with $w_{\rm eff}=w_\phi=-1$ and $\Om_\phi=1$ only for $n<0$. For $n>0$ the only stable point is the point $C$ which can not give rise to late time acceleration. This point, in fact, can not give any viable cosmological solution as it is an attarctor solution also in the case of power law potential~\cite{Bahamonde:2017ize}. Which means, for $n>0$, apparently, we do not have any late time acceleration solution. But that is not the case as from the fixed point analysis we do not have the full dynamics of the system. This can be considered as a limitation of the dynamical system analysis. This will be clearer in the next section where we discuss the dynamics with the numerical results by analysing the nature of the potentials as well as the corresponding functions of $\lam$ and $\Gamma$.

\section{Numerical study of the dynamics}
\label{sec:num}

In this section we discuss the dynamics of the scalar field qualitatively and point out the missing information about the dynamics in the fixed point analysis. 

\subsection{Nature of axion-like and power law potentials}
\label{sec:nature}
\begin{figure}[h]
\centering
\includegraphics[scale=.68]{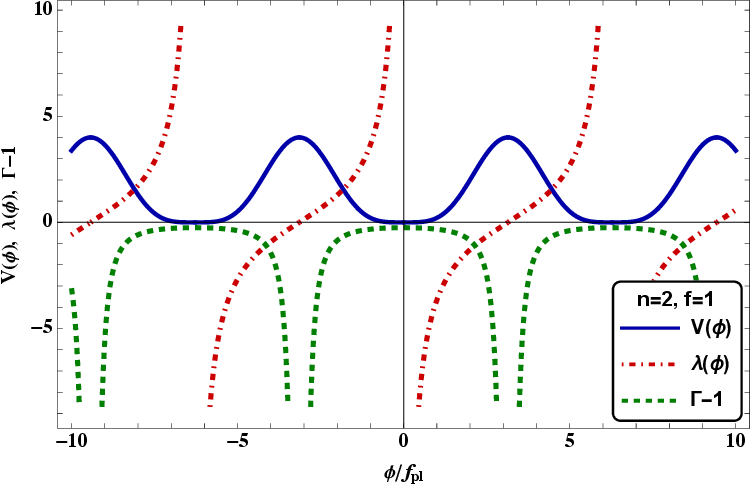} \vskip15pt
\includegraphics[scale=.68]{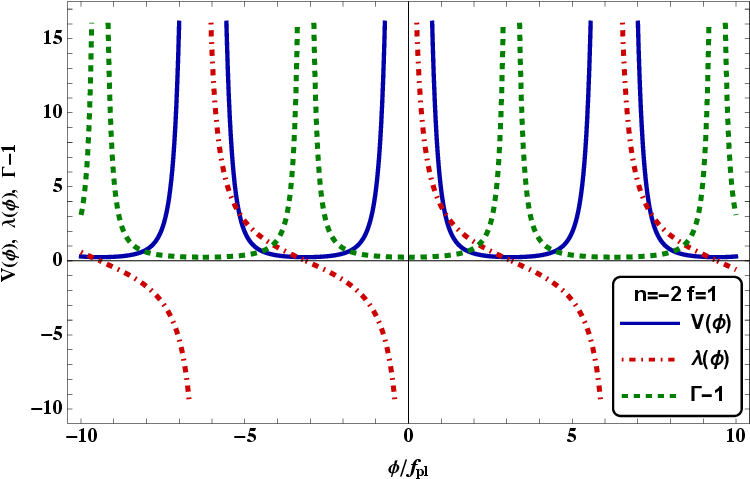}
\caption{Blue (solid), red (dot-dashed) and green (dashed) lines represent the potential~\eqref{eq:pot} normalised by $V_0$, its slope~\eqref{eq:lam} and corresponding $\Gamma-1$~\eqref{eq:Gamma}, respectively for $n=2$ (upper) and $n=-2$ (lower) with $f=1$.} 
\label{fig:pot}
\end{figure}

\begin{figure}[h]
\centering
\includegraphics[scale=.65]{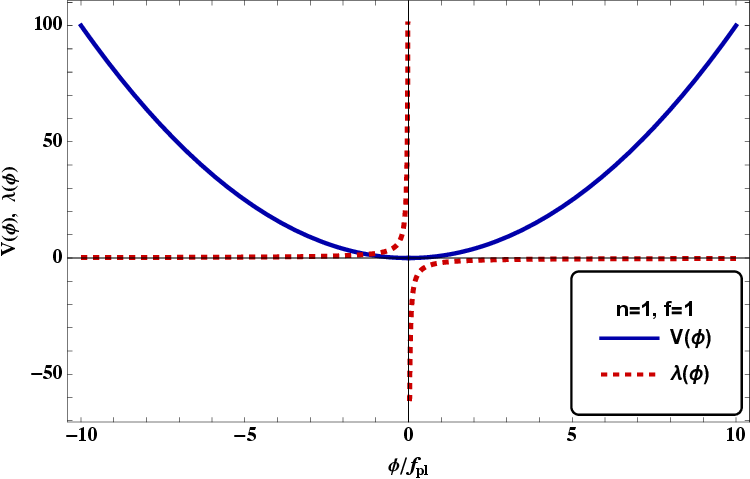} \vskip15pt
\includegraphics[scale=.65]{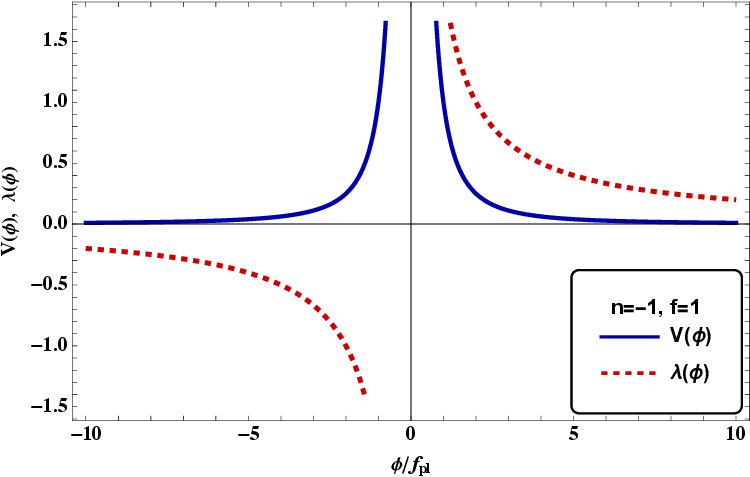}
\caption{Blue (solid) and red (dashed) lines represent the potential~\eqref{eq:pot_pl} normalised by $V_0$ and corresponding slope $\lam$ \eqref{eq:lam0_pl} for $n=1$ (upper) and $n=-1$ (lower) with $f=1$.} 
\label{fig:pot_pl}
\end{figure}

In Fig.~\ref{fig:pot} we have shown the nature of the axion-like potential~\eqref{eq:pot} and its corresponding slope $\lam$~\eqref{eq:lam0} and $\Gamma-1$~\eqref{eq:Gamma} for $n=2$ (upper) and $n=-2$ (lower). Fig.~\ref{fig:pot_pl} shows the nature of the power law potential~\eqref{eq:pot_pl} and its slope $\lam$ for both positive and negative values of $n$. For axion-like potential (Fig.~\ref{fig:pot}), while the potential is continuous for positive values of $n$ it has periodic discontinuities for negative values of $n$. But within two discontinuities the potential is continuous for negative values of $n$ which can give us interesting cosmological solutions. Now, note that, for axion-like potential \eqref{eq:pot}, the points of maxima of the potential with positive values of $n$ become the points of minima for negative values of $n$ and the points of minima for the positive values of $n$ become the points of discontinuity for negative values of $n$. So, for both the cases the potential has minima. Now, any minima at $\phi=\phi_0$ can be shifted to the origin by the transformation $\phi\to \phi\pm\phi_0$. For positive values of $n$, around the minima we can approximate the axion-like potential~\eqref{eq:pot} as the power law potential~\eqref{eq:pot_pl}.
If we compare the upper figures of Fig.~\ref{fig:pot} and \ref{fig:pot_pl} we can see that in one cycle, around $\phi=0$, the axion-like potential~\eqref{eq:pot} has similar behaviour as the power law potential~\eqref{eq:pot_pl} including the nature of their slopes. For $\phi>0$, in the half cycle of axion-like potential~\eqref{eq:pot}, the slope varies from zero (at the maxima) to $-\infty$ (at $\phi=0$) which is similar to the power potential for $\phi>0$. The same similarity is there in the nature of the slopes for $\phi<0$. So, even though these two potentials, apparently, seem different as the axion-like potential~\eqref{eq:pot} is periodic in nature but the power law potential~\eqref{eq:pot_pl} is not, they have similarities in one cycle of the axion-like potential for positive values of $n$. These similarities in the nature lead to the similarity in the cosmological dynamics of the scalar field in both the potentials. 

On the other hand, for negative values of $n$, the nature of the potentials and the corresponding functions of $\lam$ and $\Gamma$ are different. From the lower figure of Fig.~\ref{fig:pot} we can see that the axion-like potential has minima but from the lower figure of Fig.~\ref{fig:pot_pl} we see that the power law potential is a runaway potential. These figures also show the difference in the slope of the potentials. But the important difference comes in the form of the function $\Gamma$ as it becomes very large near the minima and becomes smaller as we go away from the minima for the axion-like potential (lower figure of Fig.~\ref{fig:pot}) while it is a constant for power law potential (Eq.~\eqref{eq:Gamma_pl}). This gives a major difference between the two potentials and we shall see that this nature of the function $\Gamma$ can lead to viable tracker solutions for axion-like potential unlike the power law potential \cite{Zlatev:1998tr,Steinhardt:1999nw}. In this regard, the slope of the potential also has an important role to paly which we shall also discuss.

Apart from achieving the tracker solution,for negative values of $n$, another important thing to notice is that the minima of the axion-like potential~\eqref{eq:pot}, unlike the positive $n$ case, is non zero. This plays an important role to achieve a CC like behaviour of the potential during the late time. The value of the potential at the minima would be $V_{\rm min}=2^n V_0$. Now, to achieve CC like behaviour of the potential during late time we can equate $V_{\rm min}^{1/4}$ with dark energy scale $\rho_{\rm DE}^{1/4}$, where $\rho_{\rm DE}$ is the dark energy density which gives
\begin{eqnarray}
    V_0\approx\frac{\rho_{\rm DE}}{2^n}\, ,~~{\rm with}~~n<0  \, .
    \label{eq:DE_tracker}
\end{eqnarray}
From the above equation we can see that once we fix negative values of $n$ the scale of the axion-like potential~\eqref{eq:pot} $V_0$ can be chosen suitably to achieve proper dark energy scale. Here, we should mention that for the power law potential~\eqref{eq:pot_pl}, with negative values of $n$, we can not have CC like behaviour for physical values of $\phi$. So, the existence of CC like behaviour in the axion-like potential~\eqref{eq:pot} can make the potential~\eqref{eq:pot} cosmologically viable along with tracker behaviour. 
\subsection{Positive $n$ and scaling solutions}
\label{sec:posn}

\begin{figure}[h]
\centering
\includegraphics[scale=.84]{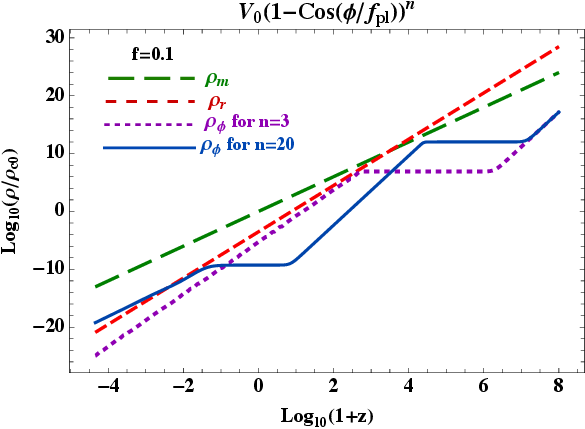} \vskip15pt
\includegraphics[scale=.84]{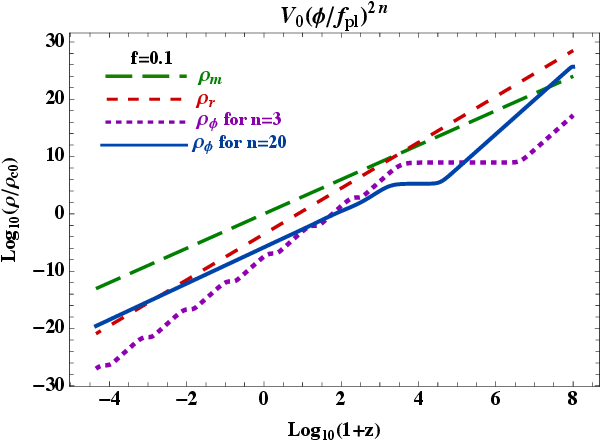}
\caption{Matter (long dashed green), radiation (short dashed red) and scalar field (solid blue (for $n=3$) and dashed purple (for $n=20$)) energy densities, normalised with the present value of critical density $\rho_{\rm c0}$, are shown for axion-like potential~\eqref{eq:pot} (left) and power law potential~\eqref{eq:pot_pl} (lower). Same initial conditions have been considered by fixing the initial value of the field at $\phi_i/\Mpl=\pi f-0.004$ which is very close to the point of maxima for axion-like potential~\eqref{eq:pot} and $\dot\phi_i/\Mpl=10^{-5}$ with $V_0=10^{6}\rho_{\rm c0}$ and $f=0.1$.} 
\label{fig:rho_n4n20f2}
\end{figure}

In both the potentials, for positive values of $n$, the scalar field will oscillate around the minima. During this oscillation we can show that for a canonical scalar field $<\frac{1}{2}\dot\phi^2>=n<V(\phi)>$, where the symbol $<\dots>$ represents average value over one cycle \cite{Turner:1983he,Johnson:2008se}. So, the average EoS of the scalar field becomes 
\begin{equation}
    <w_\phi>=\frac{n-1}{n+1}\, .
    \label{eq:w_osc}
\end{equation}
For, $n=1$ we have $<w_\phi>=0$, {\it i.e.}, $\rho_\phi\sim a^{-3}$ like matter while for $n=2$ we have $<w_\phi>=1/3$ giving rise to $\rho_\phi\sim a^{-4}$ like radiation. When $n\ll 1$ we can have $<w_\phi>\approx -1$ which may give rise to accelerated phase of the universe. So, we can have different scalar field dynamics depending upon the values of $n$. 

The function $\Gamma$, which tells us about the nature of the potential, plays an important role to determine the scalar field dynamics in the cosmological background. $\Gamma=1$ for exponential potential with constant slope which gives rise to scaling solution for $\lam>\sqrt{3}$ \cite{Copeland:1997et}. For Tracker dynamics we need $\Gamma>1$ \cite{Steinhardt:1999nw}. From the upper figure of Fig.~\ref{fig:pot} we see that, for positive $n$, around the minima of the potential, $|\Gamma-1|$ is also minimum and as we increase the value of $n$ it gets closer to zero around the minima which can be understood from Eq.~\eqref{eq:Gamma}. So, as we increase the value of positive $n$ there is a possibility to get scaling solution near the minima as in this case $\Gamma$ will be close to 1 and $\lam$ is very large which means the potential can behave like a steep exponential one giving rise to scaling behaviour. This same argument is also valid for the power law potential~\eqref{eq:pot_pl}. From Eq.~\eqref{eq:Gamma_pl} we see that $\Gamma\approx 1$ for large positive values of $n$ and from the upper figure of Fig.~\ref{fig:pot_pl} we see that the slope $\lam$ is large around the minima. So, also in power law potential we can get scaling solutions for larger positive values of $n$. In Fig.~\ref{fig:rho_n4n20f2} we have shown the evolution of the scalar field energy density $\rho_\phi$ along with matter and radiation energy densities for $n=3$ (dashed purple line) and $n=20$ (solid blue line) for both the potentials (upper one for axion-like and lower one for power law potential). We can see the similarity in the evolution of $\rho_\phi$ in both the potentials for $n=3$. This can easily be understood from Eq.~\eqref{eq:Gamma} which tells us that when $\lam$ is very large the function $\Gamma$ becomes same for both the potentials. So, the axion-like potential basically reduces to a power law potential and gives rise to similar dynamics. We can also see that during the late times $\rho_\phi$ decays faster than radiation which follows Eq.~\eqref{eq:w_osc}.

We have already argued that larger positive values of $n$ give rise to scaling solutions. Here, the solid blue line of Fig.~\ref{fig:rho_n4n20f2}, for $n=20$, shows the scaling behaviour of the scalar field dynamics for both the potentials with same initial conditions. In both the figures, $\rho_\phi$ follows the matter during lower redshifts which gives $w_\phi\approx 0$. This behaviour is an attractor solution. This tells us that the behaviour of the scalar field dynamics during the lower redshifts is same in both the potentials except that the power law potential reaches the attractor solution earlier than the axion-like potential for same initial conditions. The reason behind this can be understood from Eqs.~\eqref{eq:Gamma} and \eqref{eq:Gamma_pl} which tell us that the difference between the function $\Gamma$ of the two potentials is the last term of Eq.~\eqref{eq:Gamma}. For finite values of $n$, this term is negligible only when $\lam\to\infty$. As around the maxima of axion-like potential $\lam$ is very small and the last term of Eq.~\eqref{eq:Gamma} remains non-negligible which makes $\Gamma<1$ {\it i.e.} away from behaving like exponential potential which means away from scaling solutions. So, if we fix the initial conditions around the maxima, which is the case for Fig.~\ref{fig:rho_n4n20f2}, the scalar field will take some time to reach the scaling solutions until the slope $\lam$ becomes sufficiently large. Until then it will follow the dynamics which will be sensitive to the initial conditions. And, if we consider very large values of $n$ {\it e.g.}, $n=20$, before reaching the scaling solutions, $\rho_\phi\sim a^{-6}$ which makes $\rho_\phi$ much smaller than the background energy density very fast which causes large Hubble damping. Because of this damping the scalar field freezes to evolve causing it to behave like a CC. This CC like behaviour increases its energy density and it again starts evolving. Now, if $\lam$ is very large and $\Gamma\neq 1$, $\rho_\phi$ can repeat this behaviour and this is what is happening in the evolution of $\rho_\phi$ in the upper figure of Fig.~\ref{fig:rho_n4n20f2} for $n=20$ before reaching the attractor scaling solutions, {\it i.e.}, at the higher redshifts. This difference in the dynamics at higher redshifts comes only for very large values of $n$. For $n=3$, Fig.~\ref{fig:rho_n4n20f2} shows that the dynamics is very similar in both the potentials. Also, we can reduce this difference even for large values of $n$ by choosing suitable initial conditions.
These similarities in the dynamics suggest us that rather than the actual functional form of the potential the form and the values of the functions $\lam$ and $\Gamma$ determine the dynamics of the scalar field irrespective of the potential. So, there can be degeneracy in the dynamics of the scalar field for different potentials as we have seen in this case.

\begin{figure}[h]
\centering
\includegraphics[scale=.85]{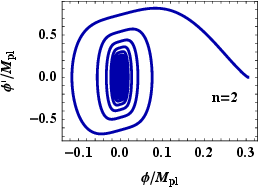} ~~~~~
\includegraphics[scale=.85]{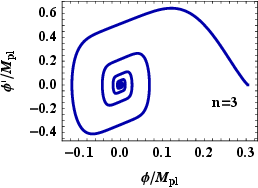} \vskip5pt
\includegraphics[scale=.85]{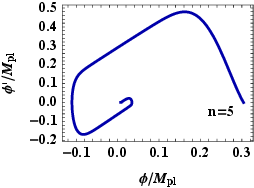}~~~~~
\includegraphics[scale=.85]{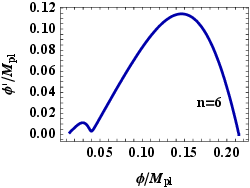}
\caption{Trajectories in the $\phi-\phi'$ plane are shown for the axion-like potential with $n=2,\;3,\; 5\; {\rm and}\; 6$. Same initial conditions ($\phi_i=(\pi f-0.1)\Mpl$ and $\phi'=10^{-5}\Mpl$), $f=0.1$ and $V_0=10^6\rho_{\rm c0}$ have been considered for all the figures.}
\label{fig:phi}
\end{figure}

Here, we have another interesting thing to notice that the Eq.~\eqref{eq:w_osc} suggests that for large positive values of $n$, $w_\phi\to 1$. But, for $n=20$, instead of having $w_\phi\approx1$ we have $w_\phi\approx 0$ which does not follow Eq.~\eqref{eq:w_osc}. This suggests that for larger positive values of $n$ the oscillatory nature of the potentials may not be valid.  In fact, numerically, we have seen that as we increase the positive values of $n$ the scalar field EoS follows the Eq.~\eqref{eq:w_osc} and increases towards 1 initially. But after some particular value of $n$, $w_\phi$ starts returning towards value $w_\phi=0$. So, there is a turning point in $w_\phi$ and we have seen that for $n>3$, $w_\phi$ starts returning towards 0. In Fig.~\ref{fig:phi} we have shown the trajectories in the $\phi-\phi'$ plane, where {\it prime} represents derivative w.r.t. $\ln(1+z)$, for increasing values of $n$, from $n=2$ to $n=6$ for axion-like potential. We can see that as we increase the values of $n$ the oscillations reduce and the oscillations seizes to exist for around $n>5$. This clearly shows that for larger values of $n$ the axion-like or power law potential becomes non-oscillatory. In Ref.~\cite{Agrawal:2019lmo} also this nature of turnaround in $w_\phi$ has been shown for power law potential but this nature is also present in the axion-like potential. The physical reason behind this is that for larger positive values of $n$ the minima of both the potentials become wider which makes it less oscillatory.

\subsection{Negative $n$ and tracker solutions}
\label{sec:negn}

\begin{figure}[t]
\centering
\includegraphics[scale=0.82]{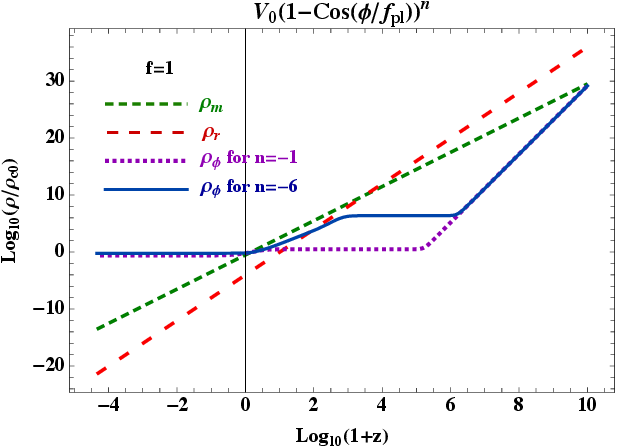} \vskip15pt
\includegraphics[scale=0.82]{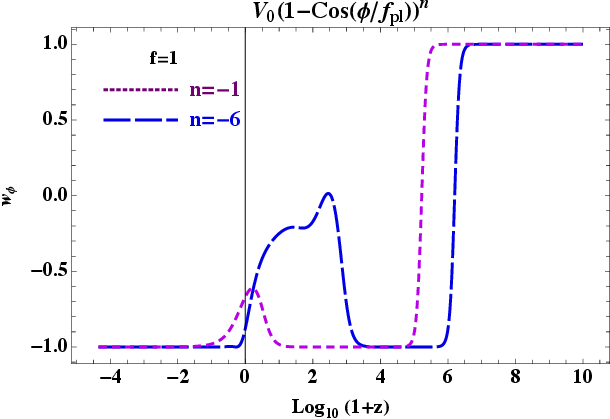}
\caption{(Left:) Matter (short dashed green), radiation (long dashed red) and scalar field (dotted purple (for $n=-1$) and solid blue (for $n=-6$)) energy densities, normalised with the present value of critical density $\rho_{\rm c0}$, are shown. (lower:) Scalar field EoS is shown for $n=-1$ (dotted purple) and $n=-6$ (dashed blue). Initial conditions are fixed by considering the present value of matter density $\Om_{m0}\approx0.3$ for $f=1$. Both the figures are for axion-like potential~\eqref{eq:pot}.}
\label{fig:cos_track}
\end{figure}

\begin{figure}[h]
\centering
\includegraphics[scale=1.2]{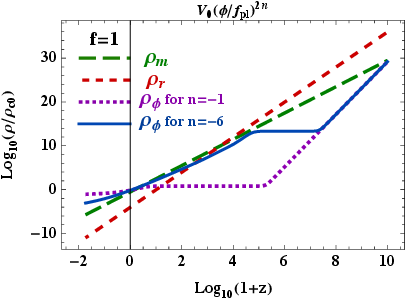} \vskip15pt
\includegraphics[scale=1.2]{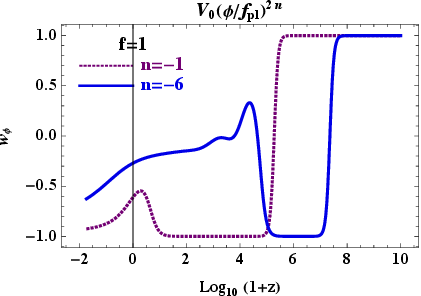}
\caption{(Left:) Matter (long dashed green), radiation (short dashed red) and scalar field (dotted purple (for $n=-1$) and solid blue (for $n=-6$)) energy densities, normalised with the present value of critical density $\rho_{\rm c0}$, are shown. (lower:) Scalar field EoS is shown for $n=-1$ (dotted purple) and $n=-6$ (solid blue). Initial conditions are fixed by considering the present value of matter density $\Om_{m0}\approx0.3$ and $f=1$, $V_0=1.6\rho_{\rm c0}$ and $5\times 10^9\rho_{\rm c0}$ for $n=-1$ and $-6$ respectively. Inital conditions are $\phi_i=0.5\Mpl$ and $\phi_i'=10^{-3}\Mpl$. Both the figures are for power law potential~\eqref{eq:pot_pl}.}
\label{fig:pl_track}
\end{figure}

While we get similar behaviour in the cosmological dynamics of the scalar field in the axion-like \eqref{eq:pot} and power law \eqref{eq:pot_pl} potentials for positive values of $n$, we shall see, as we discuss further, for negative values of $n$ the dynamics is very distinguishable in terms of the tracker solutions and existence of CC like behaviour. In fact, the axion-like potential has the ability to make the tracker solutions more viable than the power law potential as the axion-like potential~\eqref{eq:pot} exhibits CC like behaviour during late time for negative values of $n$ as discussed above by fixing the scale of the potential $V_0$ by the dark energy scale using the Eq.~\eqref{eq:DE_tracker}. 
For negative values of $n$, from Eqs.~\eqref{eq:Gamma} and \eqref{eq:Gamma_pl}, we can see that $\Gamma-1$ will be symmetrically opposite to the one for positive $n$ for both the potentials. From the lower figure of Fig.~\eqref{fig:pot} we can see that, for axion-like potential, $\Gamma-1$ is minimum near the point of discontinuity and becomes larger as we approach the minima of the potential where the slope $\lam$ is also very small. This nature in the functions $\lam$ and $\Gamma$ can give rise to interesting dynamics of the scalar field. As the scalar field rolls down the potential it rolls down from larger slope to much smaller slope towards the minima which can result to late time acceleration. At the same time, $\Gamma$ increases and eventually it becomes $>1$ which can give rise to tracker behavior \cite{Steinhardt:1999nw,Zlatev:1998tr}. Initially, while rolling down, the value of $\Gamma-1$ was small and if it is close to zero the potential will behave like an exponential potential with large slope giving rise to scaling-like behaviour. Towards the bottom of the potential since $\Gamma-1$ becomes large and $\lam$ becomes small the scaling-like behaviour is followed by a slow roll of scalar field which may result to late time acceleration. As $\Gamma-1$ eventually increases, in this case, we can not get a perfect scaling behaviour rather we get tracker behaviour in the scalar field dynamics where the scalar field almost scales the background in the past and eventually exits from the scaling-like nature and takes over matter in the recent past (upper figure of Fig.~\ref{fig:cos_track}). Figs.~\ref{fig:cos_track} and \ref{fig:pl_track} show the evolution of the energy densities and scalar field EoS for axion-like and power law potentials respectively for $n=-1$ and $n=-6$ with $f=1$ and $\Om_{\m0}\approx 0.3$. From both the figures we can see that the scalar field dynamics possesses the tracker behaviour for $n=-6$.

\begin{figure}[t]
\centering
\includegraphics[scale=1.1]{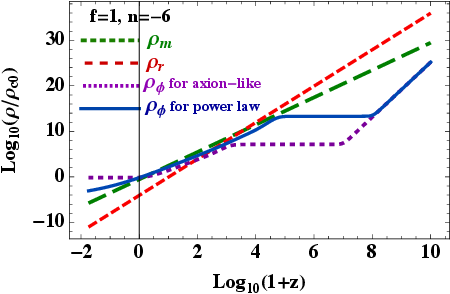} \vskip15pt
\includegraphics[scale=1.1]{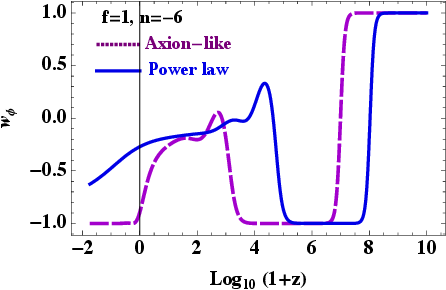}
\caption{(Top:) Matter (short dashed green), radiation (long dashed red) and scalar field (dotted purple (for axion-like potential~\eqref{eq:pot}) and solid blue (for power law potential~\eqref{eq:pot_pl})) energy densities, normalised with the present value of critical density $\rho_{\rm c0}$, are shown. (lower:) Scalar field EoS is shown for axion-like potential~\eqref{eq:pot} (dashed purple) and power law potential~\eqref{eq:pot_pl} (solid blue). Initial conditions are $\phi_i=0.5\Mpl$ and $\phi'_i=10^{-5}\Mpl$ for both the potentials and fixed by considering the present value of matter density $\Om_{m0}\approx0.3$ with $f=1$ and $n=-6$, $V_0=44.79\rho_{\rm c0}$ and $5\times10^9\rho_{\rm c0}$ for axion-like and power law potentials respectively.}
\label{fig:track}
\end{figure}
\begin{figure}[t]
\centering
\includegraphics[scale=1.1]{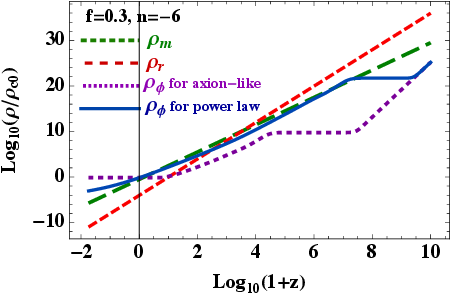} \vskip15pt
\includegraphics[scale=1.1]{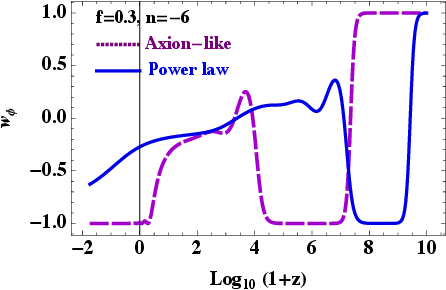}
\caption{(Top:) Matter (short dashed green), radiation (long dashed red) and scalar field (dotted purple (for axion-like potential~\eqref{eq:pot}) and solid blue (for power law potential~\eqref{eq:pot_pl})) energy densities, normalised with the present value of critical density $\rho_{\rm c0}$, are shown. (lower:) Scalar field EoS is shown for axion-like potential~\eqref{eq:pot} (dashed purple) and power law potential~\eqref{eq:pot_pl} (solid blue). Initial conditions are $\phi_i=0.1\Mpl$ and $\phi'_i=10^{-5}\Mpl$ for both the potentials and fixed by considering the present value of matter density $\Om_{m0}\approx0.3$ with $f=0.3$, $n=-6$ and $V_0=44.79\rho_{\rm c0}$ and $10^{16}\rho_{\rm c0}$ for axion-like and power law potential respectively.}
\label{fig:track_f}
\end{figure}

\noindent
For $n=-1$ we don't have tracker solutions but the dynamics is more close to thawing dynamics \cite{Caldwell:2005tm,Scherrer:2007pu} which gives rise to viable cosmology but can be very sensitive to initial conditions.  From Fig.~\ref{fig:cos_track} we see that as we increase the negative values of $n$ the tracker dynamics becomes more prevalent {\it e.g.}, for $n=-6$, same as the power law potential \cite{Zlatev:1998tr} (Fig.~\ref{fig:pl_track}). But from the lower figure of Fig.~\ref{fig:pl_track} we see that for larger negative values of $n$ the present value of scalar field EoS ($w_{\phi 0}$) becomes larger than $-1$ which can be disfavoured by the data \cite{Planck:2018vyg}. On the other hand, from the lower figure of Fig.~\ref{fig:cos_track}, we can see that for axion-like potential we get $w_{\phi 0}\approx -1$ even for larger negative values of $n$ as $\lam$ becomes very small around the minima and at some point it becomes close to zero while for power law potential the slope becomes negligibly small only asymptotically.
So, while for the power law potential the tracker solutions may not be viable for the axion-like potential the scenario changes drastically which has been depicted in Fig.~\ref{fig:track}. Fig.~\ref{fig:track} compares the tracker solution in both the potentials for $n=-6$ and $f=1$. Upper figure of Fig.~\ref{fig:track} shows the evolution of the energy densities which shows the tracker behavior of the scalar field in both the potentials. The lower figure of Fig.~\ref{fig:track} shows the evolution of the scalar field EoS in both the potentials. It can easily be seen from the lower figure that the EoS of the scalar field is close the $-1$ at present for axion-like potential while it is away from $-1$ for power law potential. If we increase the value of $n$, for axion-like potential we can still get $w_\phi$ close to $-1$ at present but for the power law potential the difference with $-1$ increases even further. The reason behind this is the existence of CC like behaviour in the axion-like potential~\eqref{eq:pot} for negative values of $n$ which can be achieved during the recent past depending upon the values of the parameter $f$. E.g., from the lower figure of Fig.~\eqref{fig:cos_track} we can see that $w_\phi\approx -0.9$ at present and becomes $-1$ in future for $f=1$ and $n=-6$. In this case we reach the CC like behaviour in future. But for lower values of $f$ this CC like behaviour can be reached in the recent past as depicted in Fig.~\ref{fig:track_f}. Fig.~\ref{fig:track_f} compares the tracker behaviour in the axion-like and power law potentials for same exponent with $f=0.3$ with same initial conditions. From the lower figure of Fig.~\ref{fig:track_f} we can see that the axion-like potential can have CC like behaviour from the recent past which makes $w_\phi$ very close to $-1$ which can not be achieved in power law potential. In this regard, we should also note that the initial value of $\phi$ should be fixed away from the minima as otherwise we can not have any nontrivial or interesting dynamics. 

So, for negative values of $n$ we have different behaviour in the scalar field dynamics which makes the two potentials distinguishable by making the tracker dynamics more cosmologically viable for the axion-like potential.

\section{Early dark energy}
\label{sec:EDE}

\begin{figure}[t]
\centering
\includegraphics[scale=.84]{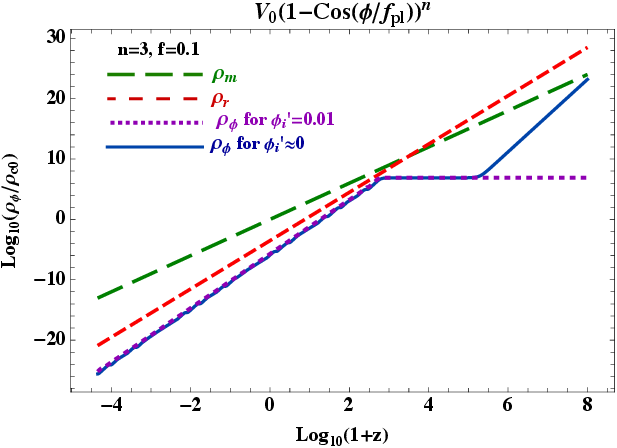} \vskip15pt
\includegraphics[scale=.84]{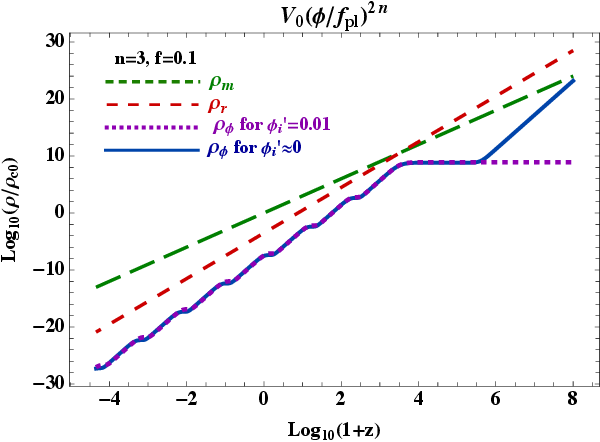}
\caption{Matter (long dashed green), radiation (short dashed red) and scalar field (solid blue (for $\phi_i'=0.01$) and dotted purple (for $\phi_i'\approx 0$)) energy densities, normalised with the present value of critical density $\rho_{\rm c0}$, are shown for the axion-like potential~\eqref{eq:pot} (upper) and power law potential~\eqref{eq:pot_pl} (lower). Both the figures are for $\phi_i=\pi f-0.01$, $n=3$, $f=0.1$ and $V_0=10^6\rho_{\rm c0}$. In the figure $\phi_i$ and $\phi_i'$ are normalised with $\Mpl$.}
\label{fig:cos_EDE}
\end{figure}

\begin{figure}[h]
\centering
\includegraphics[scale=.76]{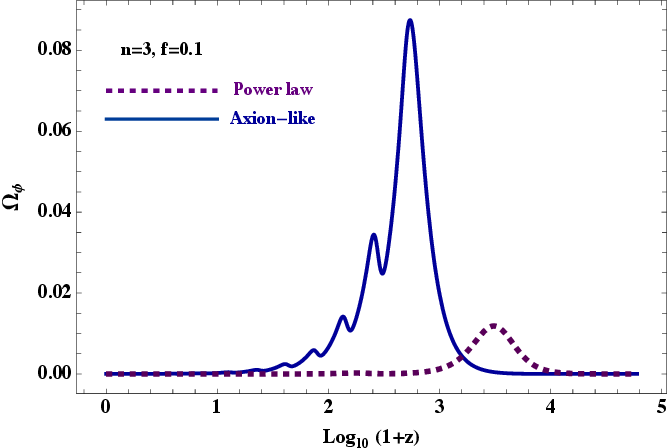} \vskip15pt
\includegraphics[scale=.76]{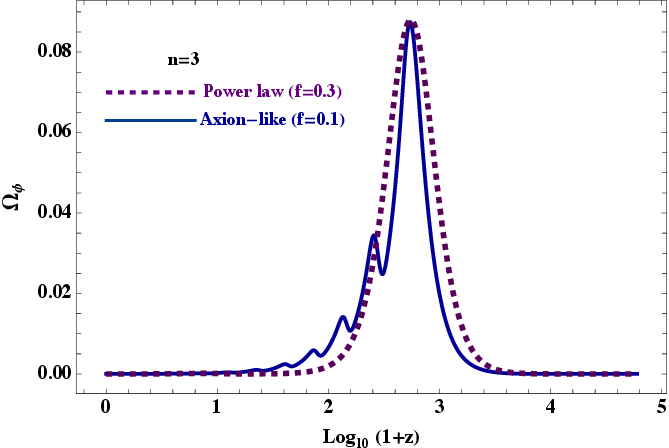}
\caption{scalar field density parameter for axion-like potential (solid blue) and power law potential (dashed purple) are shown. upper figure is for the same initial values of $\phi_i=\pi f-0.01$  and $\phi_i'\approx 0$ with $V_0=10^6\rho_{\rm c0}$. lower figure is for $V_0=2.5\times 10^4\rho_{\rm c0}$ and $f=0.3$ with $\phi=\pi f-0.01$ for the power law potential while these values are same as the upper figure for the axion-like potential.}
\label{fig:EDE_den_03}
\end{figure}

EDE solutions can alleviate $H_0$ tension to some extent \cite{Poulin:2018cxd,Poulin:2018dzj,Poulin:2023lkg,Agrawal:2019lmo,Kodama:2023fjq,Lin:2019qug,Niedermann:2019olb,Niedermann:2020dwg,Berghaus:2019cls,Berghaus:2022cwf,Karwal:2021vpk} by injecting a fraction of dark energy around the matter-radiation equality (MReq). This can be achieved with scalar fields if the scalar field freezes to evolve before the MReq which leads to a CC like behaviour in the scalar field dynamics and increases the value of $\rho_\phi$. Now, for viable cosmology, the scalar field can't remain frozen as this will result to cosmic acceleration much earlier. So, $\rho_\phi$ has to decay and to avoid interfering  with the structure formation it has to decay fast around MReq which has been shown in the Fig.~\ref{fig:cos_EDE}. This nature in the scalar field dynamics can be seen in steep potentials \cite{Geng:2015fla} for which the scalar field eventually exhibits attractor scaling behaviour as discussed above. In fact, the specific requirements in the EDE dynamics can be possible with scaling-freezing dynamics \cite{Ramadan:2023ivw,Copeland:2023zqz}.

How EDE can be useful to alleviate the Hubble tension can be understood from the definition of the angle subtended by the comoving sound horizon $r_{\rm s}$ at the decoupling epoch $z_\star$ and it is given by 
\begin{eqnarray}
    \theta_{\rm s}=\frac{r_{\rm s}(z_\star)}{D_{\rm A}(z_\star)} \, .
\end{eqnarray}
$\theta_{\rm s}$ is precisely measured by the CMB observations \cite{Planck:2018vyg,Planck:2013pxb}. In the last equation $D_{\rm A}(z_\star)$ is the comoving angular diameter distance to the CMB surface of last scatter and $r_{\rm s}(z_\star)$ and $D_{\rm A}(z_\star)$ are given by
\begin{eqnarray}
    r_s(z_\star) &=& \int_{z_\star}^{\infty} \frac{c_s(z)}{H(z)}{\rm d}z = \frac{1}{H_\star}\int_{z_\star}^{\infty} \frac{c_s(z)}{H(z)/H_\star}{\rm d}z \, , 
    \label{eq:rs}\\
    D_A(z_\star) &=& c\int_{0}^{z_\star}\frac{dz}{H(z)} =  \frac{c}{H_0}\int_{0}^{z_\star}\frac{dz}{E(z)} \, ,
     \label{eq:DA}
\end{eqnarray}
where $H_\star=H(z_\star)$, $E(z)=H(z)/H_0$, $H_0$ is the present value of the Hubble parameter, $c$ is the velocity of light and $c_{\rm s}(z)$ is the sound speed of the photon-baryon fluid which is given by
\begin{eqnarray}
    c_s(z) &=& \frac{c}{\sqrt{3(1+R)}} \, , \\
    R &=& \frac{3\omega_{\rm b}}{4\omega_\gamma}\times\frac{1}{1+z} \, , \\
    \omega_\r &=& \(1+\frac{7}{8}N_{\rm eff}\(\frac{4}{11}\)^{4/3}\)\omega_\gamma \, ,
\end{eqnarray}
where $100 {\rm km}\; {\rm s}^{-1}{\rm Mpc}^{-1}h=H_0$, $\Om_{\gamma 0}$ is photon density at present and $\omega_{\gamma} =\Om_{\gamma 0} h^2= 2.47\times10^{-5}$, $\Om_{\rm b 0}$ is the baryon density at present and $\omega_{\rm b}=\Om_{\rm b 0}h^2=0.0224\pm 0.0001$, $100\theta_{\rm s}=1.0411\pm 0.0003$ and $N_{\rm eff}=3.06$ \cite{Planck:2018vyg,Riess:2021jrx}. The limits of the integration of the Eqs.~\eqref{eq:rs} and \eqref{eq:DA} are very important as one can understand that if we modify the cosmic history only around $z_\star$ and keeps it same as standard model for $z<z_\star$ then the comoving sound horizon $r_{\rm s}$ will be changed. Now, the term within the integration in the definition of $D_{\rm A}$ will remain same as $E(z)$ does not change for $z<z_\star$. But, $\theta_{\rm s}$ is precisely measured which means it can not be changed. This forces $H_0$ to change to keep $\theta_{\rm s}$ fixed. So, by injecting EDE around $z_\star$ we can increase the expansion rate which decreases the value of $r_{\rm s}(z_\star)$ which forces $H_0$ to increase so that $\theta_{\rm s}$ remains fixed.

The axion-like and the power law potentials are two widely studied potentials as EDE solutions. EDE solutions can be achieved for the positive values of $n$ in both the potentials \cite{Poulin:2018cxd,Poulin:2018dzj,Smith:2019ihp,Agrawal:2019lmo}. As we have already argued that for positive values of $n$ the dynamics in both the potentials is almost same we expect that the EDE solutions will also be same in both the potentials.  In Fig.~\ref{fig:cos_EDE}, the EDE solutions in both the potentials have been shown for same initial conditions and parameter values. We can see the similarity in the evolution of $\rho_\phi$ in both potentials except in the values of $\Om_{\rm EDE}$ which is clear from the upper figure of Fig.~\ref{fig:EDE_den_03} where the evolution of EDE density parameter ($\Om_{\rm EDE}$) in both the scenario has been shown. Note that, $\phi=\pi f$ is one of the maxima of the axion-like potential and we have chosen the initial value of the field, $\phi_i=\pi f-0.01$ which is very close to the maxima. So, this can be nearly maximum possible difference we can have in the values of $\Om_{\rm EDE}$ for same initial conditions and parameter values. The upper figure of Fig.~\ref{fig:EDE_den_03} shows the evolution of $\Om_{\rm EDE}$ for same initial conditions and parameters and we can see that the maximum value of $\Om_{\rm EDE}$ is larger in axion-like potential which also shows the deviation of the axion-like potential from the power law behaviour. But as we have already argue that we can tune the initial conditions and the parameters to achieve similar behaviour in the $\Om_{\rm EDE}$ in both the potentials which has been shown in the lower figure of Fig.~\ref{fig:EDE_den_03}. In this figure the values of $V_0$ and $f$ are different in two potentials. While $V_0$ sets the energy scale $f$ sets the initial value of $\phi$. So, by changing the value of $V_0$ we have achieved the EDE at the same energy scale and by changing the value of $f$ we change the maximum value of $\Om_{\rm EDE}$ as $\Om_{\rm EDE,max}\approx V(\phi_{\rm c})$ where $\phi_{\rm c}$ is the field value when field is frozen. So, in both the potentials we not only have same amount of EDE but their evolution is also same. And even if it differs than that difference will have insignificant effect on the background cosmological evolution. In Ref.~\cite{Smith:2019ihp} the observational constraints in both the potentials have been studied and it is found that the constraints on the model parameters are very similar. For both the potentials $n=3$ comes out to be the best fit value for the parameter $n$ and the value of $H_0$ is $71.49$ for axion-like potential and $71.82$ for power law potential. So, we can see the results are similar which justifies our arguments.

\section{Observational Constraints on the Tracker models for Axion-like Potential}
\label{sec:obs}
As the negative values of $n$ gives interesting cosmological solutions for axion-like potential~\eqref{eq:pot} in terms of the tracker behaviour it is worth obtaining the observational constraints on the model parameters. We call the model of axion-like potential with negative $n$ as the {\bf ``Tracker-Axion"}. For the purpose of obtaining the observational constraints on the parameters we use the following data:

\noindent
{\bf Cosmic Chronometers (CC):} CC data of Hubble parameter compiled by Gomez-Valent and Amendola \cite{Gomez-Valent:2018hwc}.

\noindent
{\bf SH0ES data:} The measurement of $H_0$ by the SH0ES collaboration \cite{Riess:2021jrx}.

\noindent
{\bf Pantheon:} The distance modulus measurement of type Ia supernovae compiled in the latest Pantheon sample in terms of $E(z)$ \cite{Gomez-Valent:2018hwc}.

\noindent
{\bf CMB:} The CMB shift parameter ($R$), acoustic scale ($l_{\rm A}$) measurements and the baryon density ($\omega_{\rm b}$) from compressed
likelihood for Planck 2018 results \cite{Planck:2018vyg} obtained by Chen et al \cite{Chen:2018dbv}.

\noindent
{\bf BAO:} 6dF survey ($z=0.106$) \cite{Beutler:2011hx}, SDSS-MGS survey ($z=0.15$) \cite{Ross:2014qpa}, eBOSS quasar clustering ($z=1.52$) \cite{eBOSS:2017cqx}, anisotropic BAO measurements by BOSSDR12 at $z=0.38,\; 0.51,\; 0.61$ \cite{BOSS:2016wmc} have been considered along with the BAO measurement by BOSS-DR12 using Lyman-$\al$ samples at $z=2.4$ \cite{BOSS:2017uab}. 

\noindent
{\bf MASERS:} We have also considered angular diameter distances data measured using water megamasers under the Megamaser Cosmology Project
at redshifts $z=0.0116,\; 0.0340,\; 0.0277$ for Megamasers
UGC 3789 \cite{Reid:2008nm,Reid:2012hm}, NGC 6264 \cite{Kuo:2012hg} and NGC 5765b respectively \cite{Gao:2015tqd}.

We perform a Markov Chain Monte Carlo (MCMC) analysis using the above mentioned observational data to constrain the model
parameters. To do this we use
the publicly available code {\tt emcee} \cite{Foreman-Mackey:2012any}. To analyse the results and plot the contours of the model parameters we use the publicly available python package {\tt GetDist} \cite{Lewis:2019xzd}. Our model parameters are $\{\Om_{\m0}, r_{\rm d}, h,\; \Om_{\rm b 0}, \Om_{\r 0}, f, n\}$,  where the present Hubble parameter is given by $H_0=100{\rm km~sec^{-1} Mpc^{-1}}\times h$, the parameters $f$ and $n$ are the parameters from the choice of the potential~\eqref{eq:pot} while the sound horizon at the drag epoch $r_{\rm d}$, baryon and radiation energy densities at present, $\Om_{\rm b0}$ and $\Om_{\r 0}$, appear due to incorporation of BAO and CMB data. We have considered flat priors for the parameters and priors are given in Tab.~\ref{tab:para}.

\begin{table}[h]
\begin{center}
\caption{Prior of the parameters.}
\label{tab:para}
\begin{tabular}{c c}\hline \hline
 ~~~~~~ Parameter ~~~~~~ & ~~~~~~~~~~~~~~ Prior ~~~~~~~~~~~~~~   \\ 
\hline \hline
 $\Om_{\m 0}$ & $[0.2,0.5]$  \\
$r_{\rm d}/$Mpc & $[130, 160]$ \\
 $h$ & $[0.5,0.8]$  \\
 $\Om_{\rm b0}$ & $[0.001,0.1]$   \\
 $\Om_{\r 0}\times 10^{5}$ & $[3,10]$   \\
 $f$ & $[0.01,0.7]$   \\
$n$ & $[-1,-15]$ \\
\hline\hline
\end{tabular}
\end{center}
\end{table}

\begin{table}[h]
\begin{center}
\caption{Observational constraints of the parameters are given along with the corresponding constraints in the $\Lambda$CDM model.}
\label{tab:cons}
\begin{tabular}{c c c}\hline \hline
 ~~Parameter~~ & ~~~~~~~~~ $\Lambda$CDM ~~~~~~~~~ & ~~~~~Tracker-Axion~~~~~  \\ 
\hline \hline
 $\Om_{\m 0}$ & $0.299\pm0.0067$ & $0.3004^{+0.0065}_{-0.0076}$ \\ 
$r_{\rm d}/$Mpc & $144.6\pm 1.6$ & $144.6\pm 1.6$ \\
 $h$ & $0.699\pm0.0063$ & $0.6982\pm 0.0076$  \\
 $\Om_{\rm b0}$ & $0.046\pm 0.00083$ & $0.046^{+0.00092}_{-0.00074}$  \\
 $\Om_{\r 0}\times 10^{5}$ & $4.914\pm 0.079$ & $4.92\pm 0.1$   \\
 $f$ & $---$ & $<0.11$   \\
$n$ & $---$ & $-7.8\pm 3.5$\\
\hline\hline
\end{tabular}
\end{center}
\end{table}

\begin{figure*}[h]
\centering
\includegraphics[scale=.5]{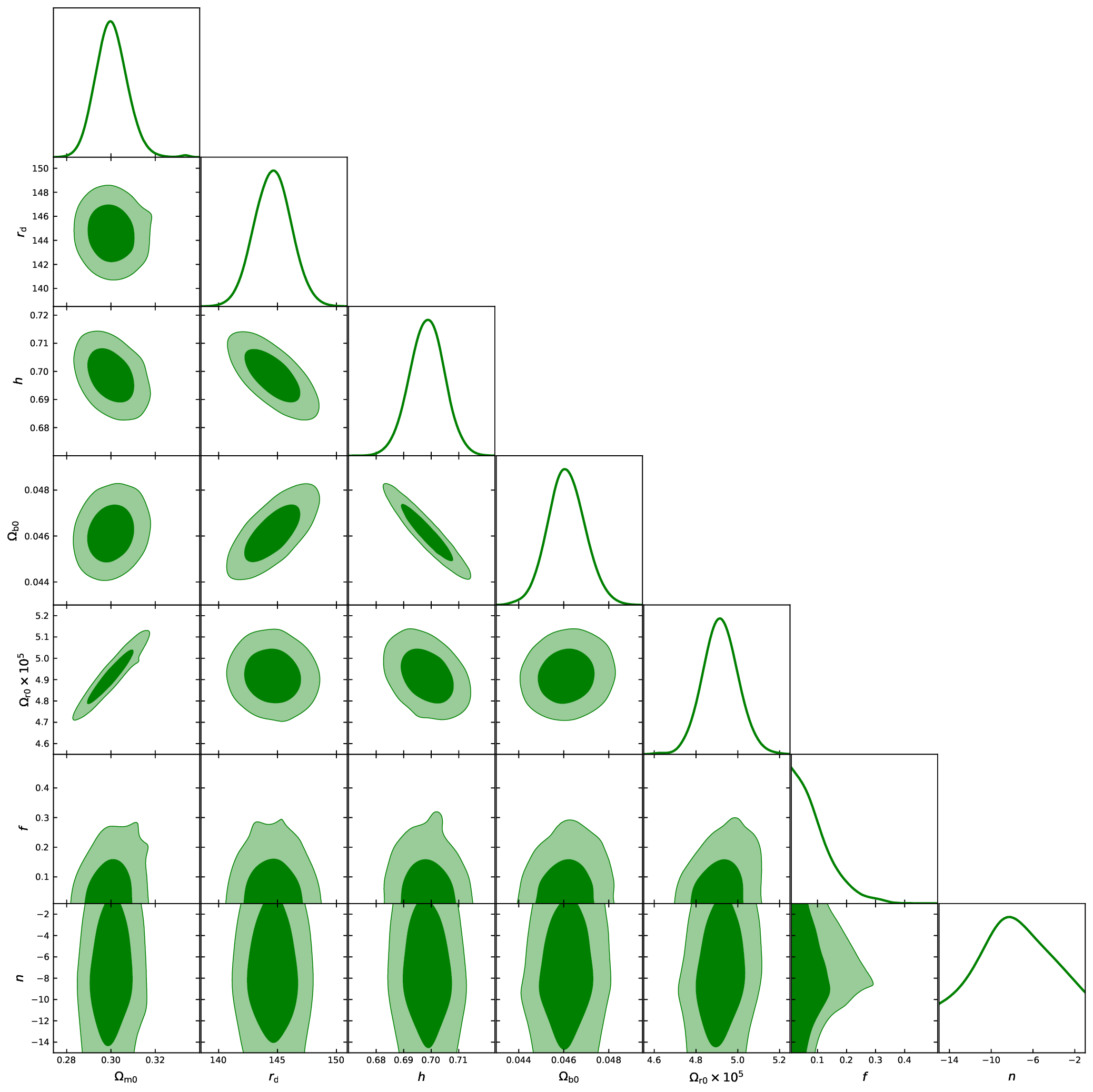}
\caption{Marginalized posterior distributions for the axion-like potential~\eqref{eq:pot}with negative $n$. The data combination is "CC$+$SH0ES$+$Pantheon$+$CMB$+$BAO$+$MASERS"}
\label{fig:obs}
\end{figure*}

The observational constraints on the parameters are given in the Tab.~\ref{tab:cons} for Tracker-Axion model. We have also included the corresponding constraints on the parameters (except $f$ and $n$) in the $\Lambda$CDM model to compare with the Tracker-Axion model. The marginalized posterior distribution of the parameters, for the Tracker-Axion model, has been shown in Fig.~\ref{fig:obs}. From Tab.~\ref{tab:cons} we see that the constraints on the parameters, in both the models, are similar. This can be understood from Fig.~\ref{fig:track_f}. From Fig.~\ref{fig:track_f} we see that the late time behaviour of the axion-field, for negative values of $n$, is very similar to CC for lower values of the parameter $f$ which has an upper bound $f<0.11$ from the considered observational data (see Tab.~\ref{tab:cons} and Fig.~\ref{fig:obs}). This nature in the EoS or in energy density of the scalar field appears because the axion-like potential~\eqref{eq:pot} with negative values of $n$ has CC like behaviour around its minima specially for $f<1$ and the scalar field reaches near the minima of the potential during the late time. In fact we can relate the dark energy scale with the axion-like potential scale $V_0$ through the Eq.~\eqref{eq:DE_tracker}. Aslo, unlike for the positive values of $n$, the axion-like potential~\eqref{eq:pot} with negative values of $n$ can not give rise to sufficient EDE as it reaches tracker behaviour and starts rolling a little before its energy density becomes sufficiently comparable to the background energy density to give rise finite amount of EDE. So, we do not expect any improvement in the present value of Hubble parameter $H_0$. Also, because of the CC like behaviour of the axion-like potential with negative values of $n$ around its minima we expect that the constraint on the cosmological parameters would be similar to $\Lambda$CDM which can be confirmed from Tab.~\ref{tab:cons}.

\section{Discussions and Conclusion}
\label{sec:conc}
In this paper, scalar field dynamics in the cosmological background has been studied for axion-like~\eqref{eq:pot} and power law~\eqref{eq:pot_pl} potentials. Both positive and negative values of the exponents have been considered. These potentials can give rise to EDE solutions and thus have the potential to reduce the $H_0$ tension \cite{Poulin:2023lkg}. This study aims to study the similarities between the two potentials in scalar field dynamics in a cosmological background. In this regard, first, we have studied the stability of the background cosmology using fixed point analysis method. The fixed points and their stability are similar for both the potentials except for the points $D_\pm$ which are hyperbolic points for axion-like potential unlike the power law potential for which these points are non-hyperbolic. So, the fixed point analysis does not give us any new information about the dynamics of the scalar field and points towards the similarity between the the axion-like and the power law potentials. This motivates us to study the scenario for both the potentials numerically. 

We then numerically studied the background cosmology for both the potentials. Possible scaling and tracker solutions, for both the potentials, have been explored. It has been found that for positive values of the exponent $n$ both the potentials can exhibit scaling solutions for larger values of $n$ and the dynamics is very similar. We do not have any distinguishable signature in any potential to differentiate. The reason behind is that, for both the potentials, for larger values of $n$, $\Gamma$ approaches towards 1 while the slope $\lam$ approaches very large values (see Figs.~\ref{fig:pot}, \ref{fig:pot_pl}). This scenario can be thought as an exponential potential with large slope ($\lam>\sqrt{3}$) which gives scaling solution. Fig.~\ref{fig:rho_n4n20f2} shows the scaling behaviour of the scalar field for the axion-like and power law potentials respectively. We can see that, for the, initial conditions they can reach the scaling regime at different time. This can be understood from the expressions of $\Gamma$. From Eq.~\eqref{eq:Gamma_pl} we see that for power law potential $\Gamma$ is close to one for large values of $n$. So, the difference is in the last term of Eq.~\eqref{eq:Gamma} which depends on $\lam$. Now in Fig.~\ref{fig:rho_n4n20f2} the scalar field's evolution is frozen twice and it freezes for the first time at high redshifts for which $n/\lam^2$ is finite. So, even for large values of $n$, $\Gamma$ may not be close to 1 at high redshifts for axion-like potentials as the last term of Eq.~\eqref{eq:Gamma} can still be finite. So, it can repeat the dynamics for steep potentials \cite{Geng:2015fla} until $\lam$ becomes very large and $\Gamma\to 1$. But by tuning the initial conditions and choosing suitable parameter values we can make $\lam$ large even at high redshifts and the dynamics of the scalar field can be almost non-distinguishable in both the potentials. So, as long as the background cosmological dynamics is concerned we can say that the axion-like and the power law potentials can give rise to similar dynamics of the scalar field for positive values of $n$. In this regard we should also mention that the two potentials have been studied as EDE solutions \cite{Poulin:2018cxd,Poulin:2018dzj,Agrawal:2019lmo} and we expect the solutions to be almost degenerate which can also be seen in ref.~\cite{Smith:2019ihp} where the authors have studied the the oscillatory potentials and the results for axion-like and power law potentials are very similar.

Unlike the positive values of the exponent $n$ for negative values of $n$ the background solutions are very distinct. In this scenario both the potentials can give rise to tracker solutions (see Figs.~\ref{fig:cos_track} and \ref{fig:pl_track}). For the power law potential, in tracker dynamics, the background cosmology may not be viable as the EoS of the scalar field, at present, becomes much larger than $-1$ (lower figure of Fig.~\ref{fig:pl_track}). But, interestingly, for axion-like potential the scenario changes drastically as we get viable cosmology with tracker solution where EoS of the scalar field is close to $-1$ at present (lower figures of Figs.~\ref{fig:cos_track}, \ref{fig:track} and \ref{fig:track_f}). This is a distinguishable feature of the axion-like potential for negative values of $n$. This discrepancy in the cosmological dynamics in the two potentials arises because of the nature of the $\Gamma$ function. From Eq.~\eqref{eq:Gamma} and \eqref{eq:Gamma_pl} we can see that the difference in these equations is in the last term of Eq.~\eqref{eq:Gamma} which plays a crucial role for negative values of $n$ but remains almost insignificant for positive values of $n$. As we can see from the lower figure of Fig.~\ref{fig:pot} that $\Gamma$ becomes much larger than one at the bottom of the potential where the slope $\lam$ becomes very small giving rise viable tracker solutions for axion-like potential. In fact, this small value of $\lam$ leads to a CC like behaviour in the axion-like potential for negative $n$ around the minima. Because of this CC like behaviour we can relate the dark energy scale with the potential scale $V_0$ using Eq.~\eqref{eq:DE_tracker}. Also, due to this CC like behaviour the observational constraint on the parameters are very similar to constraint coming on the parameters for $\Lambda$CDM model for same set of cosmological data. This can be seen from the Tab.~\ref{tab:cons} and Fig.~\ref{fig:obs}. But for power law potential $\Gamma$ remains constant and can be close to one for larger values of $n$ (Eq.~\eqref{eq:Gamma_pl}) while its slope $\lam$ \eqref{eq:lam0_pl} becomes close to zero only asymptotically. In summary, we can conclude that considering the background cosmological dynamics of the scalar field, for positive $n$, we have degeneracy among the axion-like~\eqref{eq:pot} and power law~\eqref{eq:pot_pl} potentials but the degeneracy is broken for negative values of $n$ for which we get viable tracker dynamics for axion-like potential.

\begin{acknowledgments}
The authors gratefully acknowledge the discussions regarding observational data analysis with Shahnawaz A. Adil, Sonej Alam and Sumit Kumar. MWH acknowledges the financial support from SERB, Govt of India under the Start-up Research Grant (SRG), file no: SRG/2022/002234.
\end{acknowledgments}

\bibliographystyle{apsrev4-2}
\bibliography{references}

\end{document}